\documentclass[reprint]{revtex4-1}

\usepackage{amsmath}    % need for subequations
\usepackage{graphicx}   % need for figures
\usepackage{verbatim}   % useful for program listings
\usepackage{color}      % use if color is used in text
\usepackage{hyperref}   % use for hypertext links, including those to external documents and URLs

\begin{document}

\title{Many-body exciton and inter-valley correlations in heavily electron-doped WSe$_2$ monolayers}

\author{J. Li$^{1,2}$, M. Goryca$^{1,3}$, J. Choi$^{1}$, X. Xu$^{4}$, S. A. Crooker$^{1}$}
\affiliation{$^1$National High Magnetic Field Laboratory, Los Alamos National Laboratory, Los Alamos, NM 87545}
\affiliation{$^2$Wuhan National High Magnetic Field Center, Huazhong University of Science and Technology, Hubei 430074, China}
\affiliation{$^3$Institute of Experimental Physics, Faculty of Physics, University of Warsaw, Pasteura 5, 02-093 Warsaw, Poland}
\affiliation{$^4$Department of Physics, University of Washington, Seattle, WA 98195}

%  \date{\today}

\begin{abstract}
In monolayer transition-metal dichalcogenide semiconductors, many-body correlations can manifest in optical spectra when photoexcited electron-hole pairs (excitons) are introduced into a 2D Fermi sea of mobile carriers. At low carrier densities, the formation of positively and negatively charged excitons ($X^\pm$) is well documented.  However, in WSe$_2$ monolayers, an additional absorption resonance, often called $X^{-\prime}$, emerges at high electron density.  Its origin is not understood.  Here we investigate the $X^{-\prime}$ state via polarized absorption spectroscopy of electrostatically-gated WSe$_2$ monolayers in high magnetic fields to 60~T. Field-induced filling and emptying of the lowest optically-active Landau level in the $K'$ valley causes repeated quenching of the corresponding optical absorption. Surprisingly, however, these quenchings are accompanied by absorption changes to higher-lying Landau levels in both $K'$ and $K$ valleys, which are unoccupied. These results cannot be reconciled within a single-particle picture, and demonstrate the many-body nature and inter-valley correlations of the $X^{-\prime}$ quasiparticle state. 
\end{abstract}

\maketitle

Electrostatically-doped monolayers of transition-metal dichalcogenide (TMD) semiconductors  such as WSe$_2$ and MoSe$_2$ \cite{Schaibley:2016, MakShan:2016} provide excellent platforms to study electron-electron interactions and many-body correlations, owing to their extreme 2D quantum confinement, reduced dielectric screening, and heavy carrier masses.  In conjunction with the significant improvement of material and optical quality that is provided by encapsulation in boron nitride \cite{Cadiz:2017, Ajayi:2017, Martin:2020}, the introduction of photoexcited electron-hole pairs (excitons) into a background Fermi sea of mobile carriers can lead to emergent correlated quasiparticle states that can be readily observed as discrete new resonances in optical spectra \cite{Efimkin:2017, Sidler:2016, Wang:2017NL, Scharf:2019, Rana:2020, Glazov:2020, Regan:2020, Xu:2020, Smolenski:2021}.  Thanks to the valley-specific optical selection rules in TMD monolayers \cite{Xiao:2012, Xu:2014}, correlations between particles with different spin and valley quantum numbers can be revealed using circularly polarized light. 

It is by now very well established that doping single TMD monolayers with a low density of additional holes or electrons leads to a rapid suppression of the neutral exciton absorption resonance ($X^0$), and the concomitant appearance of positively or negatively charged excitons ($X^\pm$ trions) at lower energy \cite{Urbaszek:2018, Wang:2017NL, Back:2017, Courtade:2017, Plechinger:2016, Vaclavkova:2018, Scharf:2019, Glazov:2020, Robert:2021NC}.  In the simplest picture, $X^\pm$ are quasiparticles comprising a resident electron/hole bound to the photoexcited exciton.  In the regime of small carrier densities, $X^0$ and $X^\pm$ have related descriptions as the repulsive and attractive branches, respectively, of exciton-polaron resonances that arise from the collective response of the Fermi sea to the photoexcited exciton \cite{Back:2017, Efimkin:2017, Efimkin:2021, Glazov:2020}. Similarly, $X^\pm$ can also be described in the context of four-body exciton-carrier scattering states involving excitons dressed by electron-hole pairs \textit{within} the Fermi sea \cite{Bronold:2000, Suris:2001, Suris:2003, Reichman:2019, Combescot:2018, Rana:2020, Rana:2021}; these various pictures were recently summarized by Rana \textit{et al.} \cite{Rana:2020, Rana:2021}.  

However, it has also been reported in several recent studies that an additional strong absorption feature emerges in high-quality WSe$_2$ monolayers in the regime of large electron densities \cite{Jones:2013, Wang:2017NL, Wang:2017, Scharf:2019, Barbone:2018, Li:2021, SuFeiShi:2020}. This discrete and narrow absorption resonance, often called $X^{-\prime}$, does not appear in hole-doped monolayers, does not appear in MoSe$_2$ monolayers, is energetically well below any trion or attractive-polaron resonance, and appears only when the background electron density $n \gtrsim 2\times 10^{12}$ cm$^{-2}$. Its origin is unclear, but has been variously attributed to fine structure of the $X^-$ trion, to a doubly-charged negative trion, or to a new  quasiparticle arising from exciton interactions with intervalley plasmons \cite{Jones:2013, Barbone:2018, vanTuan:2017, Scharf:2019}.  Pioneering magneto-optical studies by Wang \textit{et al.} \cite{Wang:2017} did, however, establish that the $X^{-\prime}$ resonance evolves into a series of discrete absorption peaks in applied magnetic fields $B$, which could interpreted remarkably well within a simple picture of single-particle optical transitions between Landau levels (LLs) in the valence and conduction bands. The apparent success of this single-particle picture is somewhat surprising, given the very large electron-hole Coulomb energies that exist in TMD monolayers \cite{Urbaszek:2018}, and the expectation of strong electron-electron interactions even at high background electron densities \cite{Attaccalite:2002, Donk:2018, Miserev:2019, Li:2020}.

Therefore, to further elucidate the underlying nature of the $X^{-\prime}$ resonance, we perform polarized absorption spectroscopy of electrostatically-gated WSe$_2$ monolayers in high magnetic fields to 60~T. High $B$ fields drive a repeated filling and emptying of the lowest ($0^{th}$) optically-active LL in the $K'$ valley by the dense electron sea, leading to clearly-resolved quenchings and reappearances of the corresponding optical transition due to Pauli blocking. Crucially, however, we find that these quenchings are also accompanied by changes in the energy and oscillator strength of optical absorption to higher-lying LLs in both the $K'$ and $K$ valleys, which are \textit{not} occupied. These results therefore indicate interaction physics beyond a single particle picture, and demonstrate the many-body nature and inter-valley correlations of the emergent $X^{-\prime}$ quasiparticle state. 

Figure 1(a) shows a schematic and image of a dual-gated WSe$_2$ monolayer used in these studies. The structure was assembled via dry-stacking methods and comprises a single monolayer of exfoliated WSe$_2$, sandwiched between $\sim$25~nm thick slabs of hexagonal boron nitride (hBN). Flakes of few-layer graphene (FLG) serve as the contact and top/bottom gate electrodes.  Once assembled, the structure was  placed over the core of a single-mode optical fiber. The sample-on-fiber approach avoids temperature-dependent drifts and $B$-dependent vibrations of the optical path with respect to the WSe$_2$, enabling spectroscopy in the extreme environment of pulsed magnetic fields.  White light from a Xe lamp was directed down the single-mode fiber, through the sample and a thin-film circular polarizer, and was then retro-reflected back into a multimode collection fiber. The collected light was dispersed in a spectrometer, and transmission spectra were recorded every 0.6~ms by a fast CCD camera, throughout the 50~ms duration of a 60~T pulsed magnet. Owing to the optical selection rules in monolayer TMDs \cite{Xiao:2012, Xu:2014}, absorption transitions in the $K$ and $K'$ valley (corresponding to $\sigma^+$ and $\sigma^-$ circularly-polarized light) can be separately detected in positive and negative $B$, respectively \cite{Li:2020, Mitioglu:2015, Stier:2016, Arora:2016}.  All measurements were performed in helium vapor at 4~K.  

Figure 1(b) shows how the WSe$_2$ absorption spectrum evolves with applied gate voltage $V_g$, at $B$=0. When $V_g \approx -1$V, the monolayer is at its charge neutrality point, and only absorption from the neutral exciton ($X^0$) is observed.  When doped with free holes ($V_g \leq -2$V), the $X^0$ absorption blueshifts and vanishes, and the $X^+$ trion resonance appears at lower energy. Similarly, when doped with low densities of free electrons ($0 < V_g <3$V), $X^0$ again vanishes while two new absorption lines emerge; these are the singlet (intra-valley) and triplet (inter-valley) states of negative trions ($X_{s,t}^-$) \cite{Scharf:2019, Courtade:2017, Plechinger:2016, Vaclavkova:2018, Robert:2021NC}. 

\begin{figure}[t]
\centering
\includegraphics[width=0.99\columnwidth]{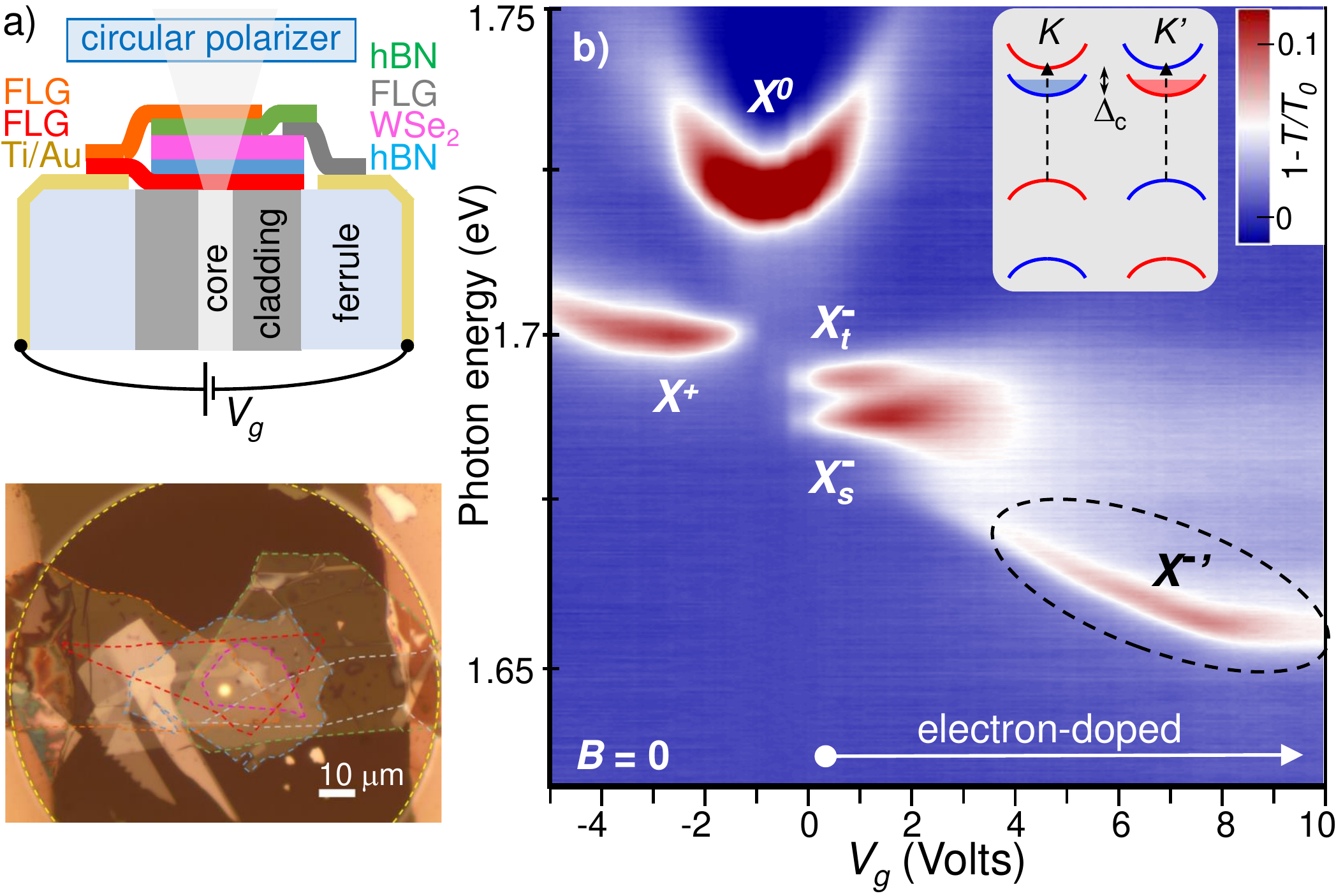}
\caption{\label{Fig1} (a) Layer schematic (top) and microscope image (bottom) of a dual-gated WSe$_2$ monolayer, which was assembled and placed over the 3.5 $\mu$m diameter core of a single-mode optical fiber. The bright spot in the image shows light emitted from the core. The dotted yellow circle shows the fiber's 125 $\mu$m diameter cladding. The assembly was mounted on a fiber-coupled probe in a $^4$He cryostat, in the bore of a 60~T pulsed magnet. (b) A map of the $B$=0 absorption spectra, $1- \frac{T}{T_0}$, at 4~K versus gate voltage $V_g$ ($T$ and $T_0$ are transmission and reference spectra). Charged exciton features ($X^+ , X^-$) appear at low hole and electron density, respectively, while the distinct $X^{-\prime}$ resonance emerges at higher electron density when $V_g \gtrsim +4$~V ($n \gtrsim 2 \times 10^{12}$ cm$^{-2}$). The kink in $X^{-\prime}$ that occurs at $V_g \approx +8$~V ($n \approx 5 \times 10^{12}$ cm$^{-2}$) corresponds to the point at which electrons begin to fill the upper CBs; \textit{i.e.}, when the Fermi energy $E_F$ equals the spin-orbit splitting of the CBs, $\Delta_c$. (Note: while such maps are more often shown with $V_g$ along the \textit{y}-axis, here we plot photon energy along the \textit{y}-axis for better comparison with the $B$-dependent maps shown in Fig. 2.)}
\end{figure}

Most important for this study, however, is the behavior at larger electron densities. When $V_g \gtrsim 4$V, both $X_{s,t}^-$ trions vanish and a pronounced additional absorption resonance appears at an even lower energy ($\sim$1.67~eV, or $\sim$15~meV below $X_s^-$).  This strong absorption line, called $X^{-\prime}$ in recent studies \cite{Wang:2017NL, Wang:2017, Li:2021}, indicates the emergence of a distinct mode having a large joint density of states with significant oscillator strength. That $X^{-\prime}$ does not appear in $n$-type MoSe$_2$, or in any $p$-type TMD monolayer, strongly suggests that it is related to the well-known conduction band (CB) structure in WSe$_2$ \cite{Kormanyos:2015, Robert:2021}. As depicted in the inset of Fig. 1(b), in each valley the optical transitions from the uppermost valence band (VB) are connected to the \textit{upper} of the two CBs. These upper CBs lie above the opposite-spin lower CBs by the spin-orbit splitting $\Delta_c$. The lower CBs do not participate in optically-allowed transitions at this energy, but they do serve as a reservoir for resident electrons, with which photogenerated excitons can interact. $X^{-\prime}$ emerges only when the lower CBs are populated by an appreciable density of electrons ($n \gtrsim 2 \times 10^{12}$~cm$^{-2}$).

\begin{figure*}[t]
\centering
\includegraphics[width=1.99\columnwidth]{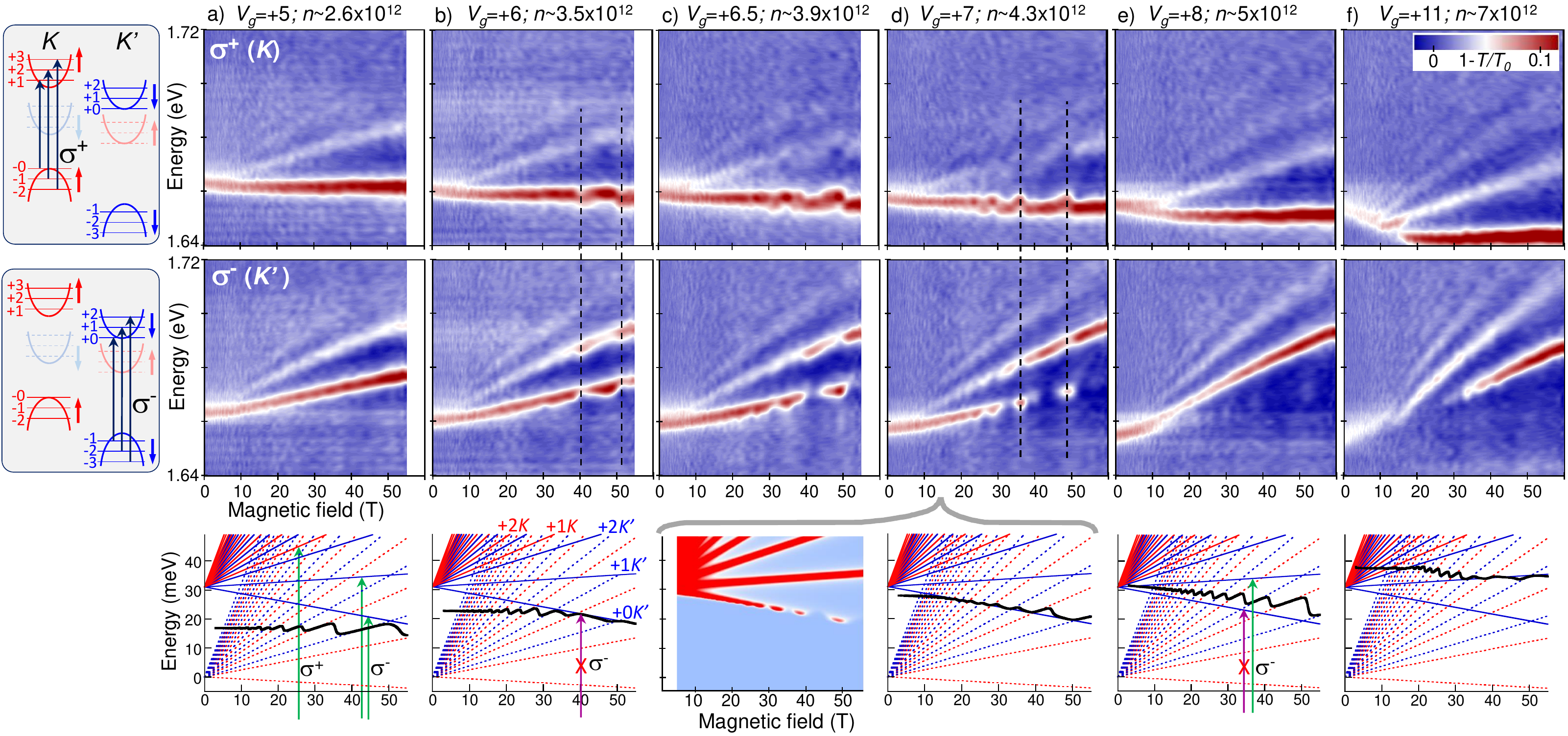}
\caption{\label{Fig2} (a-f) Color maps showing how the $X^{-\prime}$ quasiparticle resonance evolves with $B$, at electron densities $n$ from $2.6-7.0 \times 10^{12}$~cm$^{-2}$. For each $n$, the upper and lower maps show $\sigma^+$ and $\sigma^-$ absorption, respectively (i.e., optical transitions in the $K$ and $K'$ valley).  The upper and lower diagrams at the left depict the corresponding $\sigma^+$ and $\sigma^-$ allowed optical transitions between LLs in the valence and conduction bands. The bottom row shows spin- and valley-resolved LL diagrams in the CBs within a single-electron picture, along with the calculated chemical potential of the electron Fermi sea (using 1~meV LL width and $T$=4~K). Green and red vertical arrows depict allowed and Pauli-blocked optical transitions. At the smallest $n$ (panel a), the Fermi sea occupies only the lower CBs, and LLs in the upper CBs are always unoccupied (for $B<$60~T). At intermediate doping (panels b-d), electrons repeatedly fill and empty the $+0K'$ LL in the upper CB, as $B$ increases. Pauli blocking (quenching) of the corresponding $\sigma^-$ optical transition is observed whenever the $+0K'$ LL fills (see lower color maps). However, when quenching occurs, absorption to the next-higher $+1K'$ LL is \textit{enhanced}, even though this LL remains unoccupied. Moreover, absorption of $\sigma^+$ light in the $K$ valley is also modulated, despite these LLs also being unoccupied.  These correlations indicate the many-body and inter-valley correlated nature of the $X^{-\prime}$ state in $n$-type WSe$_2$ monolayers. At higher doping (panels e,f) where the Fermi sea always occupies the upper CBs, transitions to successively higher LLs are quenched.  All calculations used $\Delta_c=31$~meV, $m_e = 0.37 m_0$, $g_l = 1.2$, and $g_u=4.0$, which are in line with previously reported values \cite{Kormanyos:2015, Robert:2021}.  Using these (fixed) values, the color map in the bottom row shows the calculated available density of states in the upper CB of $K'$, for $n=4.3 \times 10^{12}$ cm$^{-2}$; repeated quenching is observed when $+0K'$ fills, but higher LLs are unaffected. Note that the overall slopes of the measured absorption lines also depend on the Zeeman splitting and cyclotron shifts of the valence bands.}
\end{figure*}

To investigate the nature of $X^{-\prime}$, we turn to polarized magneto-absorption spectroscopy in the regime of large electron density. High-field magneto-optical studies have proven to be a very effective tool for revealing how LL formation affects $X^0$ and $X^\pm$ resonances in TMD monolayers at lower carrier densities \cite{SuFeiShi:2020, Liu:2020, Li:2020, Smolenski:2019, Klein:2021}. Figure 2 shows how the $X^{-\prime}$ resonance evolves with $B$ up to 60~T, at different electron densities $n$ ranging from $2.6 - 7.0 \times 10^{12}$~cm$^{-2}$ [panels (a-f)]. At each $n$, the upper and lower colormaps show the $B$-dependent absorption of $\sigma^+$ and $\sigma^-$ circularly polarized light, which reveal the allowed optical transitions in the $K$ and $K'$ valleys, respectively. At the left of Fig. 2, the upper and lower diagrams depict the relevant absorption transitions in $K$ and $K'$, within a single-particle picture. In each valley, allowed transitions occur between LLs in the highest VB and the same-spin LLs in the upper CBs. Owing to the Berry curvature associated with massive Dirac fermions, the structure includes a 0$^{th}$ LL that is pinned to the top of the VB in the $K$ valley, and to the bottom of the CBs in the $K'$ valley \cite{Rose:2013, Cai:2013, Chu:2014}. Interband transitions are allowed between $-i \leftrightarrow i+1$ LLs (in $K$), and between $-(i+1) \leftrightarrow i$ LLs (in $K'$), where the index $i = 0, 1, 2...$ (note that different conventions are often used in the literature when labeling $K/K'$ valleys, the upper/lower CBs, and the LLs within those valleys; each study is internally consistent but care should be taken when  comparing results).

For comparison with the data, the fan diagrams shown below each pair of absorption maps depict the $B$-linear dispersion of the LLs in the upper and lower CBs (solid and dashed lines, respectively), and show how the chemical potential of the 2D Fermi sea varies within these LLs.  Red and blue lines correspond to spin-up and -down LLs. The LLs in the upper CBs have energies
\begin{align*}
E_i^{K \uparrow} &= i \hbar e B/m_e + g_u \mu_B B + \Delta_c ~ (i=1, 2, 3...) \\
E_i^{K' \downarrow} &= i \hbar e B/m_e - g_u \mu_B B + \Delta_c  ~ (i=0, 1, 2...), 
\end{align*}
while the LLs in lower CBs have energies
\begin{align*}
E_i^{K \downarrow} &= i \hbar e B/m_e + g_l \mu_B B  ~ (i=1, 2, 3...) \\
E_i^{K' \uparrow} &= i \hbar e B/m_e - g_l \mu_B B ~ (i=0, 1, 2...).  
\end{align*}
Here, $\hbar e B/m_e$ is the cyclotron energy of electrons with mass $m_e$, and upper (lower) CBs have effective \textit{g}-factor $g_u$ ($g_l$), and are split by $\Delta_c$. As shown below, these LL maps do a remarkably good job of qualitatively capturing the essential trends observed in the absorption data, despite using fixed parameter values that are not renormalized by the (varying) electron density, and using the same value of $m_e$ for both upper and lower CBs. 

Returning to the absorption maps, we first discuss spectra at smaller $n$ [see Fig. 2(a)], where $X^{-\prime}$ first emerges.  As $B$$\rightarrow$60~T the $X^{-\prime}$ resonance shifts and splits, and develops a series of additional linearly-dispersing absorption lines at higher energy, in both $\sigma^\pm$ polarizations. In good agreement with Wang \textit{et al.} \cite{Wang:2017}, these multiple lines are qualitatively consistent with a picture of free-particle transitions between LLs in the VB and in the upper CB. Within this picture (see diagrams, Fig. 2), absorption lines are separated by the sum of both electron and hole cyclotron energies, or equivalently, by an effective cyclotron energy with reduced mass $m^* = (m_e^{-1} + m_h^{-1})^{-1}$. From the data in Fig. 2(a), we determine $m^* \approx 0.28 m_0$ in $K$ and $\approx 0.31 m_0$ in $K'$, in approximate agreement with past work \cite{Wang:2017}. We note that these values are larger than $m^*$ determined from spectroscopy of neutral excitons in undoped WSe$_2$ \cite{Stier:2018}, and are almost certainly renormalized by the Fermi sea. Most importantly, however, we emphasize that at this smaller $n$, the absorption lines evolve smoothly with $B$ and do not evince any Pauli blocking effects, in accordance with the expectation that the Fermi sea resides entirely within the lower CBs (see LL fan diagram, bottom row). 

This situation changes markedly at higher electron densities where, at large $B$, the Fermi sea begins to populate LLs in the optically-active upper CB in the $K'$ valley.  When $n \approx 3.5 \times 10^{12}$ cm$^{-2}$ [see Fig. 2(b)], the evolution of the $X^{-\prime}$ resonance is only smooth until $B \approx 40$~T, at which point the $\sigma^-$ absorption to the $+0K'$ LL is abruptly quenched. At slightly larger $B$ ($\approx$45~T), the absorption reappears briefly, and is then partially quenched again above 51~T. The LL fan diagram and calculated Fermi level show that at these fields, free electrons populate, then depopulate, and then \textit{re}populate the $+0K'$ LL. Due to Pauli blocking, this modulates the absorption of $\sigma^-$ light.  Analogous quenching was also reported in recent gate-dependent studies (at fixed $B$) \cite{Wang:2017, SuFeiShi:2020}, and are in line with simple Pauli blocking of optically-active LLs. Our $B$-dependent sweeps (at fixed $n$) therefore complement earlier work, in that they allow regimes wherein a single LL can fill and empty multiple times during a single sweep, owing to the CB structure of monolayer WSe$_2$.

Crucially, however, Fig. 2(b) also reveals the presence of both inter- and intra-valley correlated behavior. Specifically, when the $+0K'$ LL fills at $\approx$40~T and 51~T and $\sigma^-$ absorption is quenched, the absorption oscillator strength to the next higher $+1K'$ LL abruptly increases. This is not expected within a picture of single-particle interband transitions because, as shown in the LL fan diagram, the $+1K'$ LL remains unoccupied for all $B$. This behavior can be understood, however, if the $X^{-\prime}$ resonance is an exciton-like quasiparticle state, which, being a bound state, has a wavefunction that is necessarily composed of many free-particle LL states in the conduction and valence bands.

Even more interestingly, the corresponding $\sigma^+$ map shows that the transition energy to the $+1K$ LL abruptly \textit{redshifts} at 40~T and at 51~T. These LLs in the opposite $K$ valley are also always unoccupied at this $n$, and therefore modulations of the transition energy to these LLs are not expected within a single-particle picture. Correspondence between $\sigma^+$ transitions in $K$ and the filling of the lowest LL in $K'$ unambiguously demonstrate that the $X^{-\prime}$ resonance is, in fact, a many-body state with inter-valley correlations.

These correlated phenomena manifest even more dramatically at larger $n$ (Figs. 2c, 2d) where the Fermi sea occupies the $+0K'$ LL over broader ranges of $B$.  Again there are clear correspondences between the occupation (quenching) and emptying (reappearance) of the $+0K'$ LL, strengthening and weakening of the oscillator strength to the $+1K'$ LL, and red/blue shifts of the $\sigma^+$ transition energy to the $+1K$ LL in the opposite valley. The color map in the bottom row shows a simple calculation of the available density of states in the upper $K'$ LLs; very good agreement with the observed quenching of the optical transition to the $+0K'$ LL is obtained using fixed values of $g_u$, $g_l$, $m_e$, and $\Delta_c$ that are in line with values reported in literature \cite{Kormanyos:2015, Robert:2021}.  We emphasize that this calculation is quite sensitive to these parameter values, and that good agreement is obtained only when $\Delta_c \approx 30$~meV. Because recent studies indicate $\Delta_c \approx 14$~meV in lightly-doped WSe$_2$ \cite{Kapu:2021}, our data suggest that $\Delta_c$ is significantly renormalized (increased) as $n$ increases, reaching $\approx$30~meV at the point where the electron Fermi sea begins to fill the upper CBs.  

At still larger $n$ (Figs. 2e, 2f) the Fermi level always resides within the upper CB.  $\sigma^-$ absorption to the $+0K'$ LL is always quenched and, with increasing $n$, absorption to higher LLs (in both valleys) is blocked.  However, as $B$ increases, these occupied LLs depopulate sequentially, leading to a series of absorption lines appearing in both $\sigma^\pm$ spectra.  The calculated position of the Fermi level within the LL fan diagrams qualitatively accounts for the reappearance of these absorption lines as $B \rightarrow$60~T.   

\begin{figure}[t]
\centering
\includegraphics[width=0.99\columnwidth]{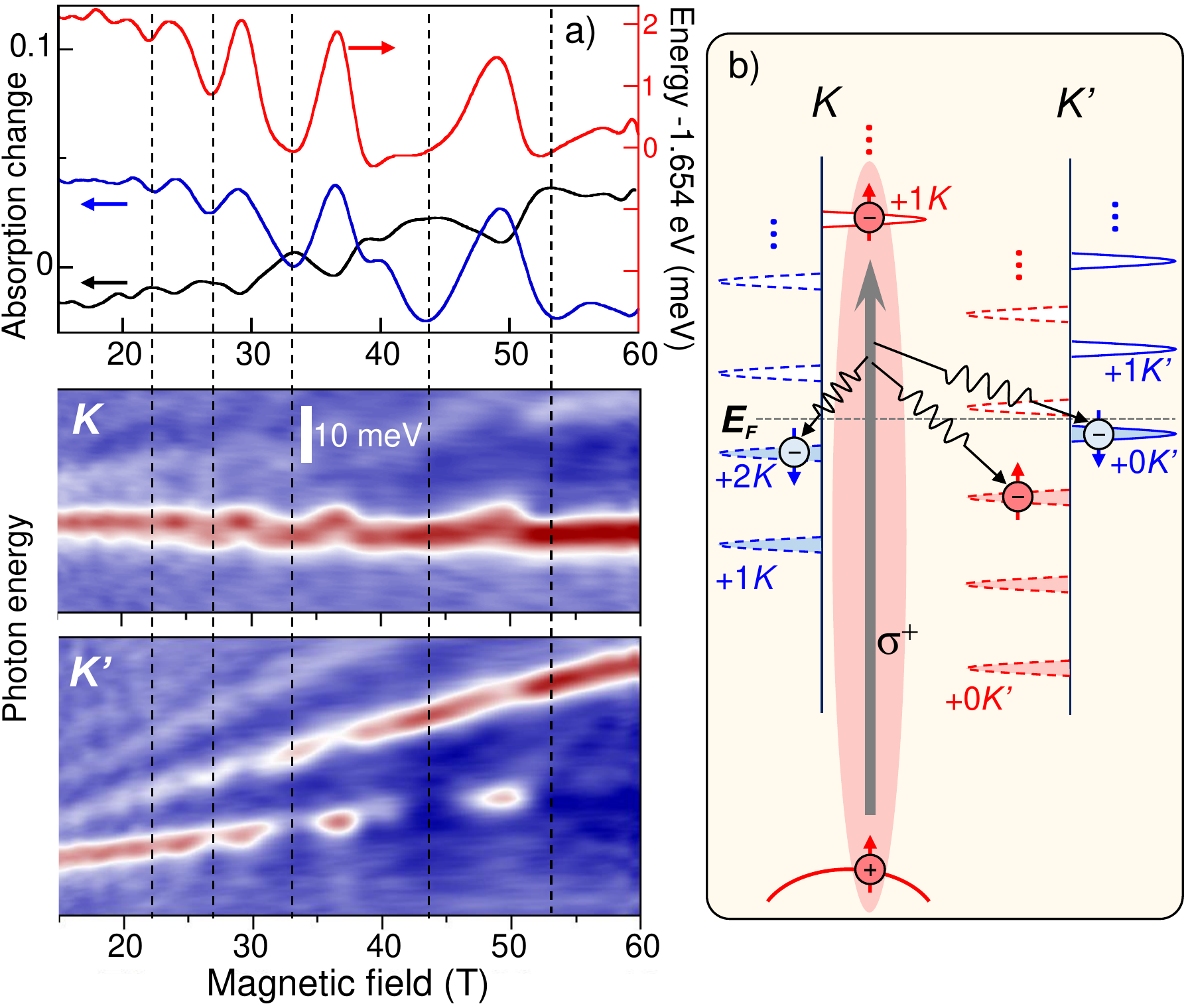}
\caption{\label{Fig3} (a) Plot showing the $B$-dependent changes in absorption of $\sigma^-$ light to the upper $+0K'$ and $+1K'$ LLs in $K'$ (blue and black curves, left axis), and how they correlate with modulations of the $\sigma^+$ transition energy to the (unoccupied) $+1K$ LL in the opposite valley (red curve, right axis), when $V_g = +7$ V.  Expanded views of the corresponding absorption maps (Fig. 2d) are shown below for reference. When $+0K'$ fills, the transition energy to $+1K$ redshifts by 2~meV, indicating that the  photoexcited spin-up exciton in $K$ becomes correlated with the reservoir of spin-down electrons in $K'$. (b) Illustration of the LLs in the upper and lower CBs in both $K$ and $K'$.  LLs in the upper and lower CBs are indicated by solid and dotted lines, respectively. Blue and red indicate spin-down and spin-up LLs. The horizontal dashed line shows the Fermi energy $E_F$, below which the LLs are filled.  This diagram depicts the situation at $\sim$33~T, where the $+0K'$ LL has just filled, which opens up an additional exciton-electron scattering channel (wavy lines) for the photoexciton in $K$.}
\end{figure}
 
Figure 3(a) quantifies the energy shift of the $\sigma^+$ optical transition to the (always unoccupied) $+1K$ LL, when $B$ forces electrons to fill and empty the $+0K'$ LL in the opposite valley. Modulations are approximately 2~meV, indicating that the photoexcited spin-up electron-hole pair in $K$ is coupled to (at least) the reservoir of spin-down electrons in $K'$. Note that the amplitude of the energy modulations increases as $B \rightarrow 60$~T, as the reservoir of spin-down electrons in $K'$ becomes more densely populated due to the increasing degeneracy of the LLs. Also shown in Fig. 3(a) for comparison are the oscillator strengths of the $\sigma^-$ transitions to the $+0K'$ LL (\textit{i.e.}, the level that is filling and emptying) and to the higher-lying $+1K'$ LL (which is always empty at this $n$). Synchronous modulations of the absorption oscillator strength to the unoccupied $+1K'$ level once again indicate that the underlying nature of these optical transitions at large electron density -- and therefore the nature of the $X^{-\prime}$ resonance -- necessarily extends beyond any single-particle picture, is excitonic in nature, and involves correlations with reservoirs of electrons having different spin/valley quantum numbers. 

We note that in the original work of Wang \textit{et al.} \cite{Wang:2017}, and also in a more recent study \cite{SuFeiShi:2020}, small discontinuities of the measured absorption lines can actually be discerned in gate-dependent sweeps (at fixed $B$), that are qualitatively consistent with our results obtained in $B$-dependent sweeps (at fixed $V_g$).  Again we emphasize the complementary nature of these two measurement approaches, which will undoubtedly help in achieving a more complete understanding of these many-body correlation effects in TMD monolayers.

Figure 3(b) depicts the relevant LLs in all four CBs, under conditions where electrons have just filled the lowest spin-down $+0K'$ level in $K'$ (\textit{e.g.}, corresponding to the experimental situation shown in Fig. 2(d), where $n \approx 4.3 \times 10^{12}$ cm$^{-2}$ and $B \approx 33$~T).  The sudden presence of spin-down electrons in $K'$ opens up a new channel with which the photoexcited spin-up electron-hole pair in $K$ can interact and scatter. This evidently lowers the energy of the corresponding optical transition (by $\sim$2~meV), analogous to the lowering of the exciton transition energy by $\sim$30~meV when neutral excitons $X^0$ first start to interact and scatter and bind with low densities of electrons to form simple $X_{s,t}^-$ charged excitons, or attractive polaron states.  Implicit in this diagram is that a new interaction (scattering) channel opens up for the spin-up exciton in $K$, whenever a sufficient density of free electrons having different spin/valley quantum numbers becomes available. In this way, the $X^{- \prime}$ absorption resonance that is observed at large $n$ in WSe$_2$ monolayers is likely best described as a many-body correlated state involving not only the photoexcited electron-hole pair, but also opposite-spin electrons in the same valley, \textit{as well as} both up- and down-spin electrons in the opposite valley.  

Whether such multi-particle correlated states appear in optical spectra depends, of course, on the structure of the CBs.  In MoSe$_2$ where $\Delta_c$ has opposite sign and optical transitions at the fundamental ``A'' exciton couple to the lower (not upper) CBs, only interactions with electrons in the lower CB in the opposite valley are possible, and therefore only a single $X^-$ trion is expected (and observed), at least until very large $n$ where the Fermi energy exceeds $|\Delta_c|$. However, optical transitions from the ``B'' exciton in MoSe$_2$, which couple to the upper CBs, are analogous to the situation observed here in WSe$_2$, and new many-body resonances appearing at moderate electron densities may appear (although improvements in optical and material quality may be necessary to spectrally resolve such shifts at the ``B'' exciton). 

In summary, polarized absorption spectroscopy in high magnetic fields demonstrates that the $X^{- \prime}$ resonance, which emerges in WSe$_2$ at high electron doping, is a many-body exciton state with inter-valley correlations. The use of high fields to 60~T allows to explore regimes where these correlations appear and disappear multiple times in a single field sweep, and to spectrally resolve the small energy shifts associated with the presence/absence of additional correlation (scattering) channels. Looking forward, analogous correlation phenomena can also be expected in highly electron-doped WS$_2$ monolayers, which have a similar CB structure to that of WSe$_2$. An especially interesting case is that of monolayer MoS$_2$, because $\Delta_c$ is very small and its sign (and therefore the ordering of the upper and lower CBs) remains a topic of some ongoing debate. Indications of a spectral feature resembling the $X^{- \prime}$ resonance have, in fact, been reported recently at larger electron densities \cite{Roch:2019, Klein:2021, Klein2:2021}, indicating that MoS$_2$ may already be a viable and additional material platform for studying related correlation phenomena in 2D semiconductor materials.

We gratefully acknowledge H. Dery and F. Rana for helpful discussions, and N.P. Wilson and A.V. Stier for technical input during initial stages of this work. Work at the NHMFL was supported by the Los Alamos LDRD program (J.L. and M.G.) and the DOE BES `Science of 100 T' program (S.A.C.). The NHMFL is supported by National Science Foundation (NSF) DMR-1644779, the State of Florida, and the U.S. Department of Energy (DOE). Work at the University of Washington was supported by the DOE BES Materials Sciences and Engineering Division DE-SC0018171. M.G. was also supported by the Norwegian Financial Mechanism 2014-2021 under Grant No. 2020/37/K/ST3/03656 and by the Polish National Agency for Academic Exchange within Polish Returns program under Grant No. PPN/PPO/2020/1/00030.

%\bibliography{MyLibrary2}{}
%\bibliographystyle{plain}

\end{document}